\documentclass[12pt,preprint]{aastex}
\usepackage{graphicx}
\begin{document}
\def\la{\mathrel{\hbox{\rlap{\hbox{\lower4pt\hbox{$\sim$}}}\hbox{$<$}}}}
\def\ga{\mathrel{\hbox{\rlap{\hbox{\lower4pt\hbox{$\sim$}}}\hbox{$>$}}}}
\def\lam{$\lambda$}
\def\kms{km~s$^{-1}$}
\def\vphot{$v_{\rm phot}$}
\def\ang{~\AA}
\def\syn{{\bf Synow}}
\def\tex{$T_{\rm exc}$}
\def\ve{$v_{\rm e}$}
\def\halpha{H$\alpha$}
\def\6200{the 6200\ang\ absorption}
\def\5700{the 5700\ang\ absorption}

\title {HYDROGEN IN TYPE~Ic SUPERNOVAE?}

\author {David Branch\altaffilmark{1},
David~J. Jeffery\altaffilmark{1}, Timothy R. Young\altaffilmark{2}, \&
E.~Baron\altaffilmark{1}}

\altaffiltext{1} {Homer L. Dodge Department of Physics and Astronomy,
University of Oklahoma, Norman,~OK 73019; e-mail: branch@nhn.ou.edu}

\altaffiltext{2} {Department of Physics, University of North Dakota,
Grand Forks, ND, 58202}

\begin{abstract}

By definition, a Type~Ic supernova (SN~Ic) does not have conspicuous
lines of hydrogen or helium in its optical spectrum.  SNe~Ic usually
are modelled in terms of the gravitational collapse of bare
carbon--oxygen cores.  We consider the possibility that the spectra of
ordinary (SN~1994I--like) SNe~Ic have been misinterpreted, and that
SNe~Ic eject hydrogen.  An absorption feature usually attributed to a
blend of Si~II \lam6355 and C~II \lam6580 may be produced by
H$\alpha$.  If SN~1994I--like SNe~Ic eject hydrogen, the possibility
that hypernova (SN~1998bw--like) SNe~Ic, some of which are associated
with gamma--ray bursts, also eject hydrogen should be considered.  The
implications of hydrogen for SN~Ic progenitors and explosion models
are briefly discussed.

\end{abstract}


\section{INTRODUCTION}

A Type~II supernova (SN~II) has conspicuous hydrogen lines in its
optical spectrum.  A Type~Ib supernova lacks conspicuous hydrogen
lines but does have conspicuous He~I lines.  A Type~Ic supernova has
conspicuous lines of neither hydrogen nor He~I.  For a review of
supernova spectral classification see Filippenko (1997).

Supernovae of these types are thought to result from core--collapse in
massive stars.  A SN~Ib is hydrogen--deficient in its outer layers,
and displays He~I lines owing to nonthermal excitation by the decay
products of radioactive $^{56}$Ni and $^{56}$Co (Lucy 1991).  In
recent years it has becomes clear that at least some SNe~Ib are not
completely hydrogen--free; in fact, it is likely that most or even all
SNe~Ib eject a small amount of hydrogen at high velocities (Deng et
al. 2000; Branch et~al. 2002; Elmhamdi et~al. 2006).  If some of the
events classified as Type~Ib had been observed earlier, hydrogen lines
might have been conspicuous, in which case they would be classified as
Type~IIb --- the designation used for events such as SN~1993J that
looked like a Type~II at early times but later looked like a Type~Ib
(Filippenko, Matheson, \& Ho 1993).

A SN~Ic either is helium--deficient in its outer layers or fails to
nonthermally excite its helium (Woosley \& Eastman 1997).  It has
become common practice to model SNe~Ic in terms of core collapse in
bare (or nearly bare) carbon--oxygen cores (Iwamoto et~al. 1994; Foley
et~al. 2003; Mazzali et~al. 2004).

Supernova spectral features are P--Cygni features characterized by
line--centered emission components and blueshifted absorption
components, with the absorptions frequently being more identifiable.
Spectral features usually are blended owing to the huge Doppler
broadening.  Nevertheless, some identifications of spectral lines are
definite.  For example, blends of Fe~II lines and features produced by
Ca~II H\&K and the Ca~II infrared triplet appear in all types of
supernovae as long as the temperature is sufficiently low.  However,
there also are some serious identification ambiguities.  The most well
known involves the Na~I D--line doublet at mean wavelength \lam5892
and the strongest optical line of He~I, \lam5876, which are separated
by only about 800 \kms.  If the corresponding observed feature is
strong and other He~I lines are not present, the feature is produced
at least mainly by Na~I, while if other He~I lines are clearly present
then the feature is at least partly due to He~I.  But if the observed
feature is not strong, it can be difficult to choose between Na~I and
He~I.

When SNe~Ic are interpreted in terms of bare carbon--oxygen cores, an
absorption feature usually near 6200\ang, which we will refer to as
\6200, is attributed to the strongest optical line of Si~II, \lam6355
(the transition definitely responsible for the similarly located deep
absorption in Type~Ia supernovae), perhaps blended with the strongest
optical line of C~II, \lam6580, forming in higher--velocity ejecta
than Si~II.  Local--thermodynamic--equilibrium (LTE) calculations of
Sobolev line optical depths for a composition dominated by carbon and
oxygen (with hydrogen and helium burned to carbon and oxygen, and
solar mass fractions of heavier elements) show that within certain
intervals of temperature Si~II \lam6355 and C~II \lam6580 are expected
to have significant optical depths (Hatano et~al. 1999).
Nevertheless, the identification of \6200\ is plagued by ambiguities.
The strongest line of Ne~I, \lam6402, is about 2200 \kms\ to the red
of Si~II \lam6355, and \halpha\ \lam6563 is only about 800 \kms\ to
the blue of C~II \lam6580. Because these four ions, each of which
could appear in supernova spectra (although Ne~I probably would
require nonthermal excitation), have their strongest optical lines to
the red, but not too far to the red, of \6200, the ambiguity is
difficult to resolve when the observed feature is weak and other lines
of these ions do not produce identifiable features.

In this paper we are primarily concerned with the possible presence of
hydrogen in SNe~Ic.  This issue (and that of helium in SNe~Ic) has
been addressed in the literature before; for summaries see Filippenko
(1997); Matheson et~al. (2001); and Branch (2002).  The presence of
hydrogen has been suggested, for example, by Filippenko (1988, 1992);
Filippenko, Porter, \& Sargent (1990); Jeffery et al. (1991); and
Branch (2002).  The currently prevailing view, however, is that
hydrogen is absent and \6200\ is produced by Si~II and/or C~II
(Wheeler et~al. 1994; Millard et~al. 1999).  In \S2, based on a
comparison of spectra of the Type~Ib or Type~Ib/c SN~1999ex, which we
believe to contain hydrogen, and the Type~Ic SN~1994I, we raise the
question of whether hydrogen also is present in SN~1994I.  In \S3,
comparisons of spectra of SN~1999ex and SN~1994I with synthetic
spectra generated with the parameterized resonant--scattering code
{\bf Synow} are presented and discussed.  In \S4, the implications of
the presence of hydrogen in ordinary (SN~1994I--like), and possibly
even in hypernova (SN~1998bw--like; Foley et~al. 2003) SNe~Ic, are
briefly considered.

\section{COMPARISON OF SN~1999ex AND SN~1994I}

SN~1999ex was initially classified as Type~Ic based on a resemblance
of its early spectrum to SN~1994I (Hamuy \& Phillips 1999), but
because it later developed He~I lines Hamuy et~al. (2002) revised the
classification to intermediate Type~Ib/c.  Although the He~I lines
were not as deep as in most events that have been classified as
Type~Ib, their presence was definite, so Branch (2002) referred to
SN~1999ex as a ``shallow--helium'' Type~Ib.  Hamuy et~al. (2002)
labelled \6200\ as Si~II, but Branch (2002) argued that the correct
identification is \halpha, consistent with the presence of \halpha\ in
most if not all other SNe~Ib.  Recently Elmhamdi et~al. (2006) have
reinforced the conclusion that SNe~Ib, including SN~1999ex and another
transition Type~Ib/c event, SN~1996aq, eject hydrogen.  (The
overluminous Type~Ib SN~1991D (Benetti et~al. 2002) could be an
exception.)

The spectra of SN~1999ex and the Type~Ic SN~1994I shown in Figures 1
to 3 are from Hamuy et~al. (2002) and Filippenko et~al. (1995),
respectively.  The horizontal scale on all figures in this paper is
logarithmic wavelength, which allows the widths of Doppler--broadened
features to be compared on an equal basis across the whole spectrum.
The spectra are ``tilted'' by multiplying by \lam$^\alpha$, with
$\alpha$ chosen to make the flux peaks near 4600\ang\ and 6300\ang\
about equally high.  The tilting makes it easier to compare spectral
features.  When comparing SN~1999ex and SN~1994I spectra we
artificially blueshift those of SN~1999ex in order to compensate for
the different photospheric velocities and roughly align the absorption
features.

In Figure~1 a spectrum of SN~1994I obtained 2 days before the time of
maximum light in the $B$ band (day~$-2$) is compared with a day~$-1$
spectrum of SN~1999ex that has been blueshifted by 4000~\kms.
Figure~2 is like Figure~1, but for day~+4 spectra of SN~1994I and
SN~1999ex, with the latter blueshifted by 2000~\kms.  Figure~3 is for
a day~+10 spectrum of SN~1994I and a day~+13 spectrum of SN~1999ex,
with the latter blueshifted by 2000~\kms.  In all three figures the
spectra are similar in many respects, although the features are more
washed out in SN~1994I because of its higher photospheric velocity at
each epoch (hence the necessity to blueshift SN~1999ex to illustrate
the similarities).  Otherwise the main differences are that He~I
\lam6678 and \lam7065, although weak, are clearly present in SN~1999ex
(making it a Type~Ib or at least a Type~Ib/c), but they are not
clearly present in SN~1994I (making it a Type~Ic).  However, it is not
possible on the basis of Figures 1 to 3 to exclude the presence of
weak He~I features in SN~1994I.  Given the similarities in these
figures --- and they are rather striking --- the question becomes: are
the Na~I and Si~II/C~II identifications correct for SN~1994I?  If so,
then the resemblance of the SN~1999ex and SN~1994I spectra from about
5500\ang\ to 6300\ang\ is coincidental.

\section{COMPARISONS WITH SYNTHETIC SPECTRA}

To further investigate the possibility of hydrogen (and He~I) in
SN~1994I we have used the parameterized resonant--scattering {\bf
Synow} code (Branch et~al. 2002) to generate synthetic spectra for
comparison with spectra of SN~1999ex and SN~1994I.

\subsection{SN 1999ex}

In Figure~4 the day~+4 spectrum of SN~1999ex is compared with a
synthetic spectrum in which \6200\ is produced by \halpha.  In the
synthetic spectrum the velocity at the photosphere is 8000 \kms.
Hydrogen is detached from the photosphere at 13,000 \kms\ where
\halpha\ has an optical depth of 0.5.  Hydrogen--line optical depths
decrease outward exponentially but very slowly, with e--folding
velocity $v_e$ = 20,000 \kms, to an imposed maximum velocity of 20,000
\kms.  Given the closeness of the fit to \6200, we know that with fine
adjustments of the optical--depth profile we could refine the fit to
be practically perfect.  Thus at the \syn\ level of analysis, \halpha\
is completely adequate and we would not prefer a different
identification without other evidence.  For all ions used in Figure~4,
reference--line optical depths and $v_e$ values (in units of 1000
\kms) are given in Table~1.

Elmhamdi et~al. (2006) show that Si~II \lam6355 cannot account for
\6200\ of SN~1999ex on its own because even when it is undetached from
the photosphere, its synthetic absorption feature is too blue.  It is
unlikely that \6200\ is due mainly to Ne~I \lam6402 because the
observed feature is strong enough that when it is fitted with Ne~I
\lam6402 other Ne~I lines appear in the synthetic spectrum and the fit
deteriorates.  This conclusion is based on LTE excitation and must be
tested against detailed non-LTE calculations that take nonthermal
excitation into account.

Hamuy et~al. (2002) also obtained a day~+4 spectrum of SN~1999ex that
extends to nearly 2.5 microns.  {\bf Synow} fitting parameters used to
fit the optical He~I lines also give satisfactory fits to features
attributed to He~I \lam10830 and He~I \lam20851 (Branch 2002; Elmhamdi
et~al. 2006), but the infrared spectrum does not provide a useful
constraint on the presence of hydrogen.

\subsection{SN 1994I}

{\bf Synow} fits to spectra of SN~1994I in which \6200\ is attributed
to Si~II and/or detached high--velocity C~II are shown in Millard
et~al. (1999) and Elmhamdi et~al. (2006).  For this paper we began
trying to fit the day~+4 spectrum of SN~1994I by varying the input
parameters that were used for the synthetic spectrum shown in Figure~4
for SN~1999ex, i.e., we assumed the presence of hydrogen and He~I
lines.  Figure~5 for SN~1994I is like Figure~4 for SN~1999ex. The
synthetic spectrum has a velocity at the photosphere of 11,000 \kms\
(compared with 8000 \kms\ for SN~1999ex).  Hydrogen lines are detached
from the photosphere at 15,000 \kms\ where \halpha\ has an optical
depth of 0.4.  Hydrogen--line optical depths decrease outward
exponentially with e--folding velocity $v_e$ = 20,000 \kms\ to an
imposed maximum velocity of 22,000 \kms.  This hydrogen optical--depth
profile is quite similar to that used for SN~1999ex in Figure~4, and
in Figure~5 \6200\ is fit nicely.  The 5700~\AA\ absorption also is
fit nicely by He~I \lam5876.  In the synthetic spectrum He~I \lam6678
and \lam7065 are weak and rather washed out owing to the high
photospheric velocity, but they do more good than harm and their
presence in the observed spectrum cannot be excluded.  (However,
regarding the strong observed absorption near one micron, the
situation remains as illustrated and discussed in Millard
et~al. (1999); the synthetic He~I \lam10830 absorption is too weak and
narrow to account entirely for the observed absorption.)
Reference--line optical depths and $v_e$ values for Figure~5 are given
in Table~1.

The synthetic spectrum in Figure~6 is like that of Figure~5 except
that it includes Si~II lines instead of hydrogen lines.  The figure
illustrates that in SN~1994I, as well as in SN~1999ex, Si~II \lam6355
is too blue to account for \6200\ on its own (flux differences between
observed and \syn\ spectra are to be expected, but differences in
wavelengths of absorption features are regarded as serious
disrepancies), although it is possible to obtain a fit by invoking
detached high--velocity C~II.  Because \6200\ is less deep in SN~1994I
than in SN~1999ex, and the SN~1994I features are more smeared out by
the higher photospheric velocity, it also is difficult to exclude Ne~I
\lam6402 in SN~1994I.

\section{DISCUSSION}

\subsection{Spectroscopy}

We have discussed the possibility of \halpha\ in absorption, in the
spectrum of the Type~Ic SN~1994I.  As shown in Figure~7, this is an
issue for SNe~Ic in general; the spectra within the wavelength
interval emphasized in this paper have strong similarities.  Other
evidence for hydrogen in SNe~Ic was emphasized by Filippenko
(1988\footnote{Filippenko (1988) referred to the events suggested to
have hydrogen as Type~Ib because the Type~Ic classification (Wheeler
\& Harkness 1986) was not yet commonly used, but these events are now
classified Type~Ic.}; 1992): the apparent presence of broad \halpha\
emission in near--maximum--light spectra.  This feature can be seen in
the day~$-2$ spectrum of SN~1994I in Figure~1 (beneath a narrow
\halpha\ emission due to an H~II region).  In order to produce a
peaked rather than a flat \halpha\ emission component the hydrogen
would have to be present down to the photosphere, rather than being
confined to a detached high--velocity shell such as we have invoked
for the absorption component.  In this regard it is interesting to
note that in Figures~1 to 3 the red edges of the putative \halpha\ and
He~I \lam5876 absorptions in SN~1994I are less sharp than in
SN~1999ex; a detached shell produces sharp red edges.  A continuous
\halpha\ optical--depth distribution that peaks at 15,000 \kms\ but
extends down to the photosphere might suffice to produce a peaked
\halpha\ emission.  This can only be tested by non-LTE calculations
because the resonance--scattering approximation on which \syn\ is
based is known to be a poor approximation for \halpha\ in SNe~II.  It
is not clear that a large hydrogen mass would be required because the
appearance of hydrogen and He~I lines depends on the radial
optical--depth profiles (Branch et~al. 2002) and on the distribution
of $^{56}$Ni (Woosley \& Eastman 1997).

At present it is difficult to establish the presence or absence of
hydrogen in SN~1994I--like SNe~Ic, but the issue should be considered
further.  A large--scale comparative study of observed spectra, like
that of Elmhamdi et~al. (2006) but including SNe~Ic as well as SNe~Ib,
probably would yield clues.  In addition, detailed non-LTE spectrum
calculations for a grid of explosion models are needed.  Such
calculations, even based on parameterized explosion models rather than
full hydrodynamical calculations, could probe several issues: e.g., is
Ne~I \lam6402 really a candidate to produce \6200, and under what
circumstances can peaked \halpha\ emission be produced?  In the only
full non-LTE calculations for SNe~Ic published so far (Baron
et~al. 1999), synthetic spectra for models containing no hydrogen did
not provide good fits to \6200\ in SN~1994I.

An interesting recent case is that of SN~2005bf (Anupama et~al. 2005;
Tominaga et~al. 2005; Folatelli et~al. 2006), which like SN~1999ex was
initially classifed as Type~Ic but later developed definite He~I
lines, making it either Type~Ib or Type~Ib/c (or Type~Icb?).  In each
of these three papers the 6200~\AA\ absorption was attributed to
\halpha.  Strong support for the identification was provided by the
presence in early spectra of Ca~II and Fe~II at the same high velocity
(15,000 \kms) as the \halpha\ absorption (Folatelli et~al. 2006).  The
very long rise time to maximum light of 40~days indicates that the
progenitor was massive.  In addition, SN~2005bf showed evidence for
high--velocity polar ejection, reminiscent of SN~1998bw--like
explosions of massive stars, some of which produce gamma--ray--bursts
(GRBs).  In view of the high ejected mass and the polar ejection,
Folatelli et~al. suggested that SN~2005bf may have been a transition
event between ordinary SNe~Ib/c and hypernovae.  Therefore if
SN~2005bf ejected hydrogen we should consider the possibility that
SN~1998bw--like SNe~Ic also do.

In SN~1998bw--like spectra \5700\ and \6200\ are assumed to be Na~I
and Si~II, respectively.  These identifications initially were adopted
(Iwamoto et~al. 1998; Branch 2001) partly because they were assumed to
be correct for SN~1994I, and SN~1998bw--like spectra were and still
are regarded to be Doppler--broadened versions of SN~1994I--like
spectra, and partly because Monte Carlo spectrum calculations for bare
carbon--oxygen cores do provide reasonable fits to SN~1998bw--like
spectra (e.g., Mazzali, Iwamoto, \& Nomoto 2000, Mazzali et~al. 2002)
with Si~II and/or C~II producing \6200.  Thus, even if \6200\ in
SN~1994I--like SNe~Ic is produced by \halpha, it may still turn out to
be produced by Si~II in SN~1998bw--likes.  Nevertheless, the
possibility that even in SN~1998bw--likes the correct identification
is \halpha\ rather than Si~II (and perhaps He~I rather than Na~I)
should be considered.  Full non--LTE spectrum caclulations for
comparison with SN~1998bw--like spectra are in their initial stages
(E.~Baron \& M.~Troxel, in preparation).

\subsection{Implications for Progenitors and Explosion Models}

Ideas concerning the progenitors of SNe~Ib and SN~1994I--like SNe~Ic
have been reviewed by Podsiadlowski (1996), who concluded that most of
the progenitors are stars of main--sequence masses $\la$20 solar
masses that lose their envelopes by means of binary interactions.  In
standard binary evolutionary scenarios, if all or even just a
substantial fraction the helium envelope must be lost, all of the
hydrogen is lost.  For example, Nomoto et~al. (1994) favored a
scenario involving two common--envelope episodes that results in a
carbon--oxygen core containing little or no helium and no hydrogen.
However, as emphasized by Podsiadlowski, there are many unresolved
questions in binary stellar evolution.  If SN~1994I--like SNe~Ic eject
hydrogen (and therefore also, of course, helium), some modification of
the standard evolutionary scenarios, perhaps involving substantial
mixing of hydrogen prior to and/or during the explosion, will be
needed.

SN~1998bw--like SNe~Ic, including those that produce GRBs, are usually
regarded to come from stars of sufficiently high main--sequence mass
that they can lose their envelopes by means of single--star winds.
Those that produce GRBs may have to be in binaries in order to have
sufficient angular momentum at the time of core--collapse for the
collapsar model (McFadyen \& Woosley 1999) to be viable (Fryer \&
Heger 2005; Petrovic et~al. 2005).  The uncertainities involved in the
evolutionary scenarios for SN~1998bw--likes are even greater than for
SN~1994I--likes, and if SN~1998bw--likes eject hydrogen, current
scenarios will require revision.  Particularly interesting in this
regard are recent results on chemically homogeneous evolution of
rapidly rotating massive stars (Yoon \& Langer 2005; Woosley \& Heger
2006).  Some of these GRB--candidate models contain hydrogen at the
time of core collapse.

We are grateful to Mario Hamuy, Alex Filippenko, and Tom Matheson for
providing spectra, to Filippenko, Alicia Soderberg, and the anonymous
referee for helpful comments on the manuscript, and to Norbert Langer
for calling our attention to recent results on quasi--homogeneous
evolution of rapidly rotating massive stars.  This work has been
supported by NSF grants AST-0204771 and AST-0506028, NASA LTSA
grant NNG04GD36G, and NASA ADP grant NAG5-3505.

\clearpage     

\begin{figure}
\includegraphics[width=.8\textwidth,angle=270]{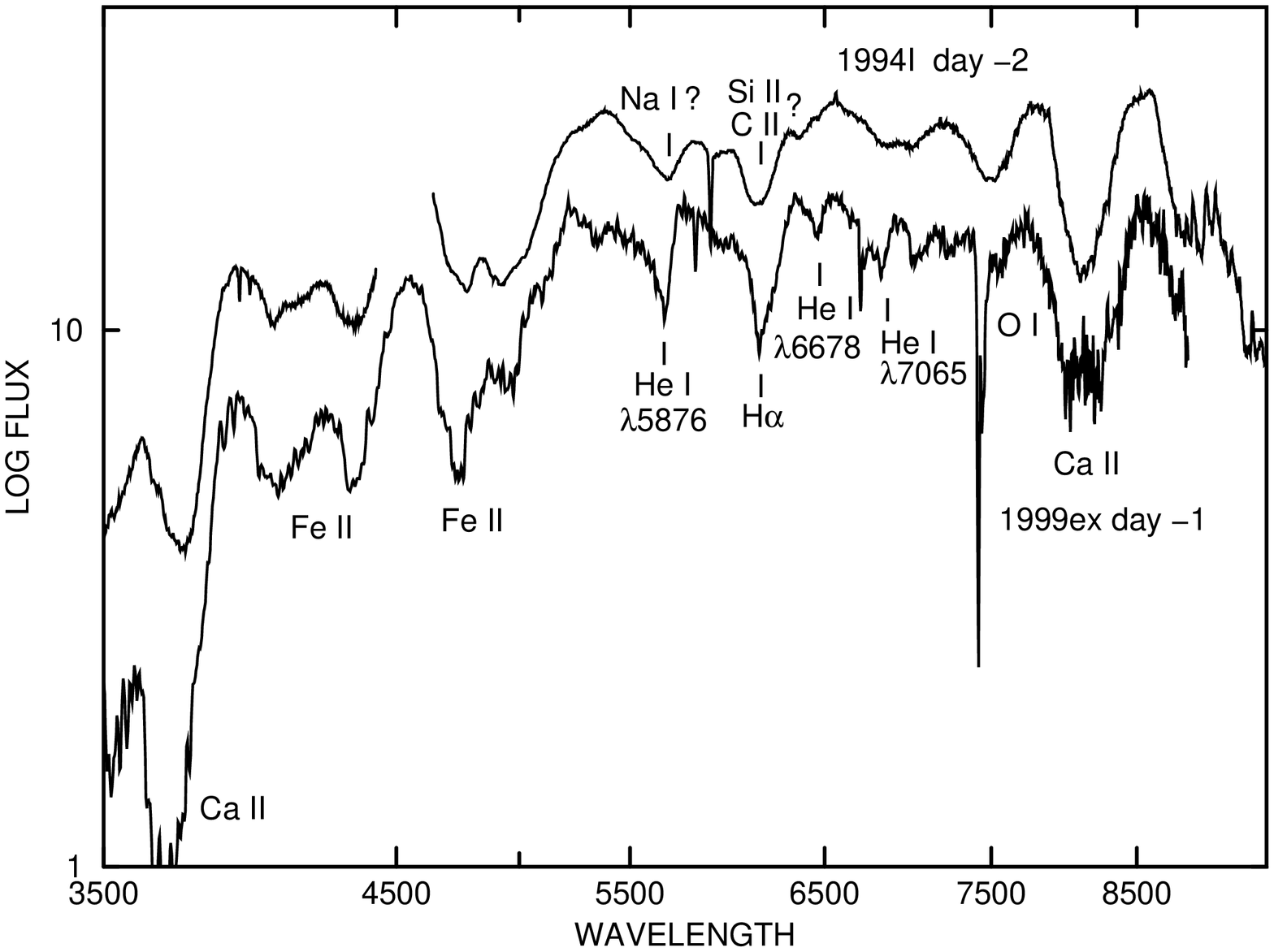}
\caption{A day~$-2$ spectrum of the Type~Ic SN~1994I is compared with
 a day~$-1$ spectrum of the Type~Ib or TypeIb/c SN~1999ex.  The latter
 is blueshifted by 4000 \kms\ to align the absorption features.  The
 flux scale is arbitrary and the spectra have been tilted to
 facilitate the comparison of spectral features.  The narrow
 absorptions near 5900\ang\ in SN~1994I and 5800\ang\ in SN~1999ex are
 interstellar Na~I in the host galaxies and the narrow absorptions
 near 6700\ang\ and 7400\ang\ in SN~1999ex are telluric.}
\end{figure}

\begin{figure}
\includegraphics[width=.8\textwidth,angle=270]{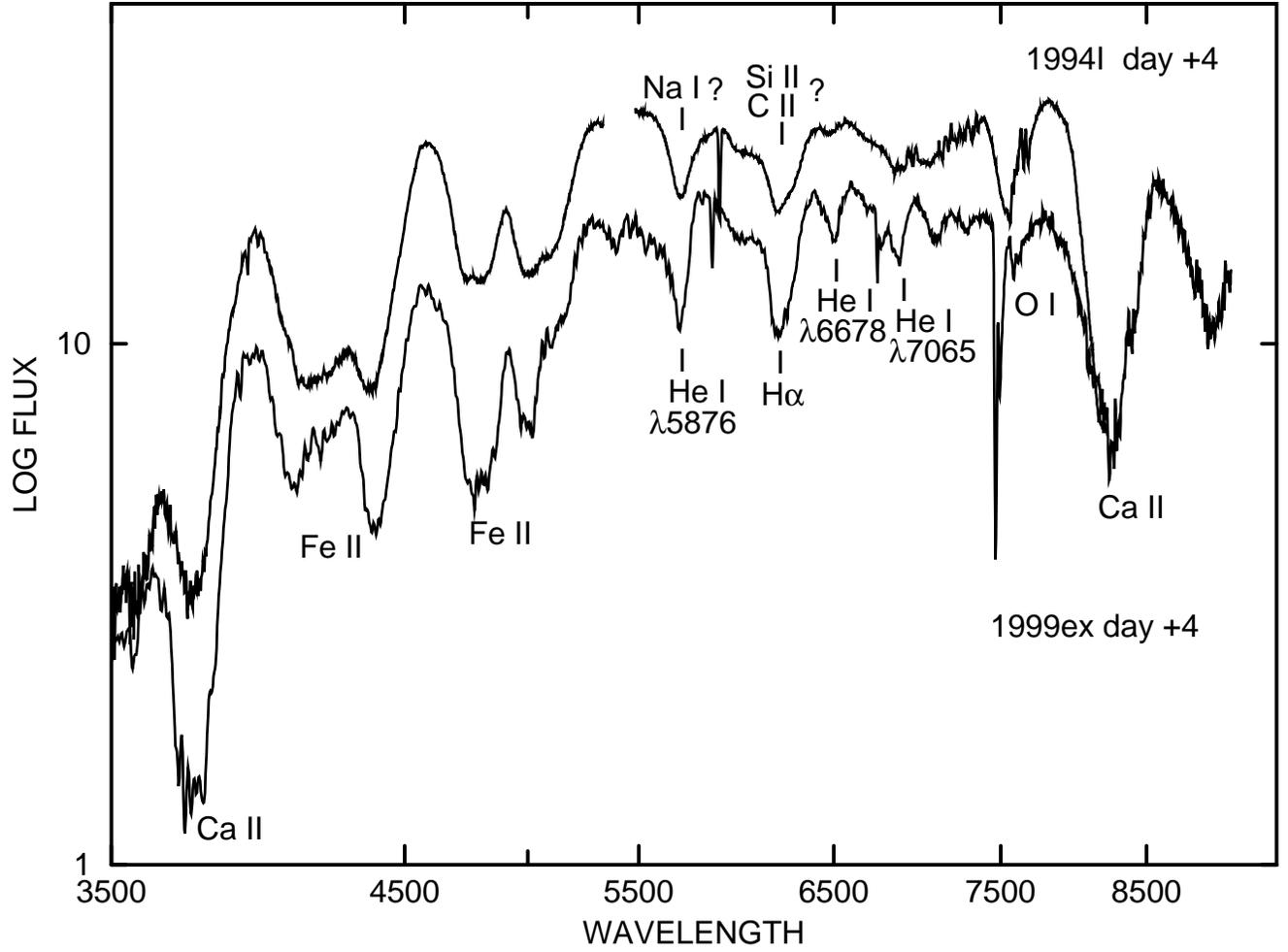}
\caption{Like Figure~1 but for day~+4 spectra of both SN~1994I and
  SN~1999ex.  The latter is blueshifted by 2000
  \kms.}
\end{figure}

\begin{figure}
\includegraphics[width=.8\textwidth,angle=270]{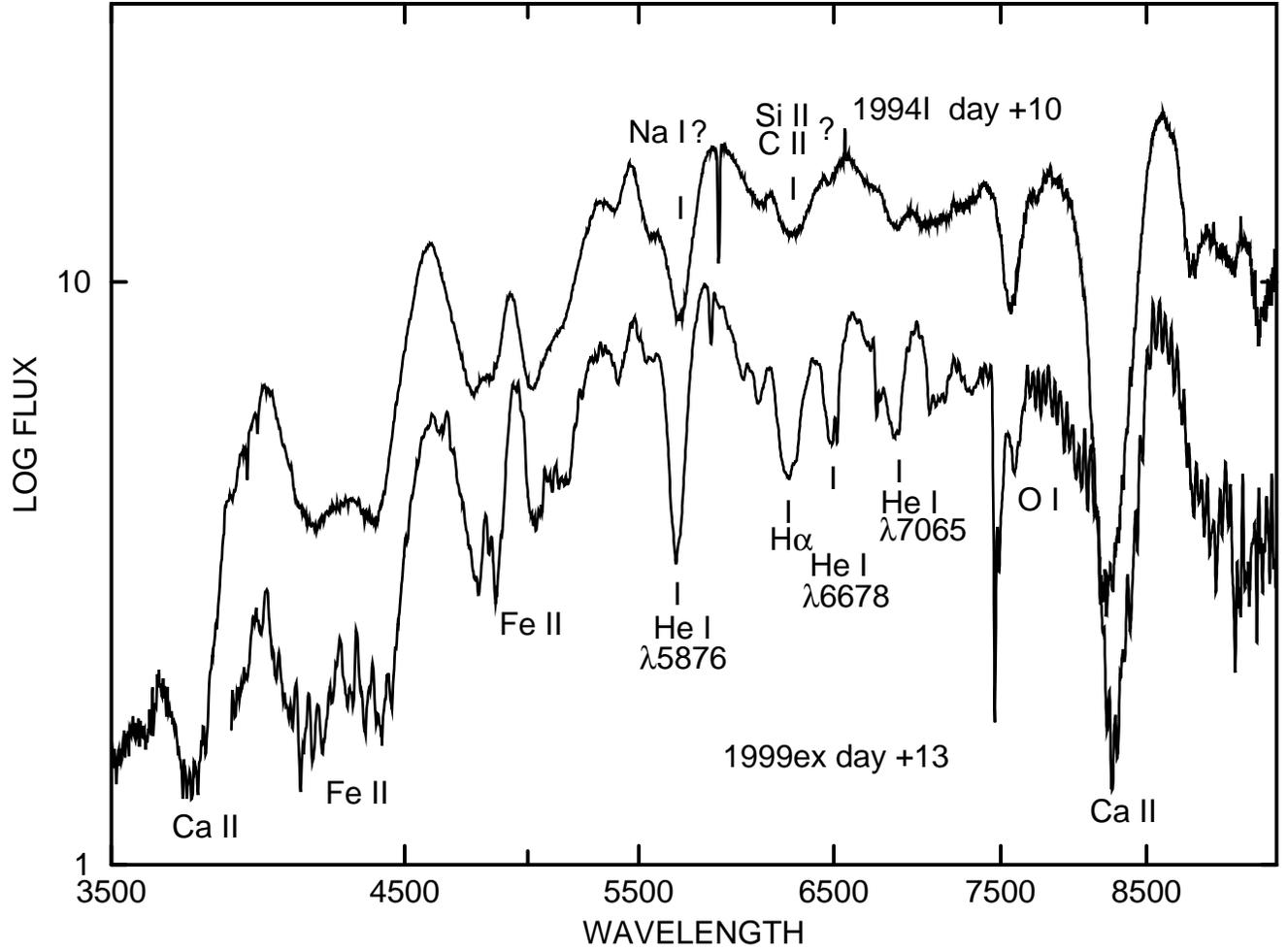}
\caption{Like Figure~1 but for a day~+10 spectrum of SN~1994I
 and a day~+13 spectrum of SN~1999ex.  The latter
 is blueshifted by 2000 \kms. }
\end{figure}

\begin{figure}
\includegraphics[width=.8\textwidth,angle=270]{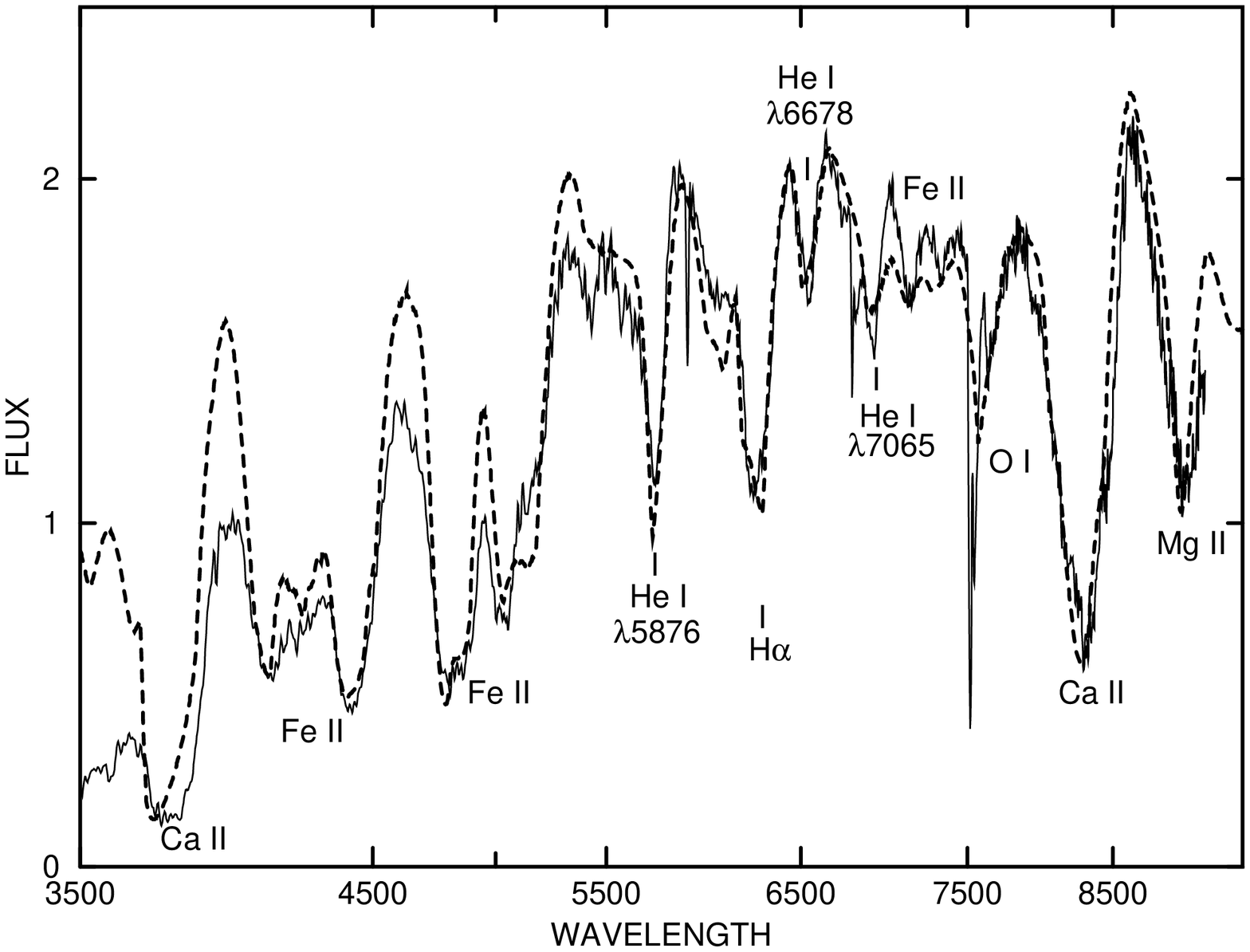}
\caption{The day~+4 spectrum of SN~1999ex ({\sl solid line}) is compared
to a {\bf Synow} synthetic spectrum ({\sl dashed line}) that has
\vphot = 8000 \kms\ and includes lines of hydrogen, He~I, O~I, Mg~II,
Ca~II, and Fe~II. }
\end{figure}

\begin{figure}
\includegraphics[width=.8\textwidth,angle=270]{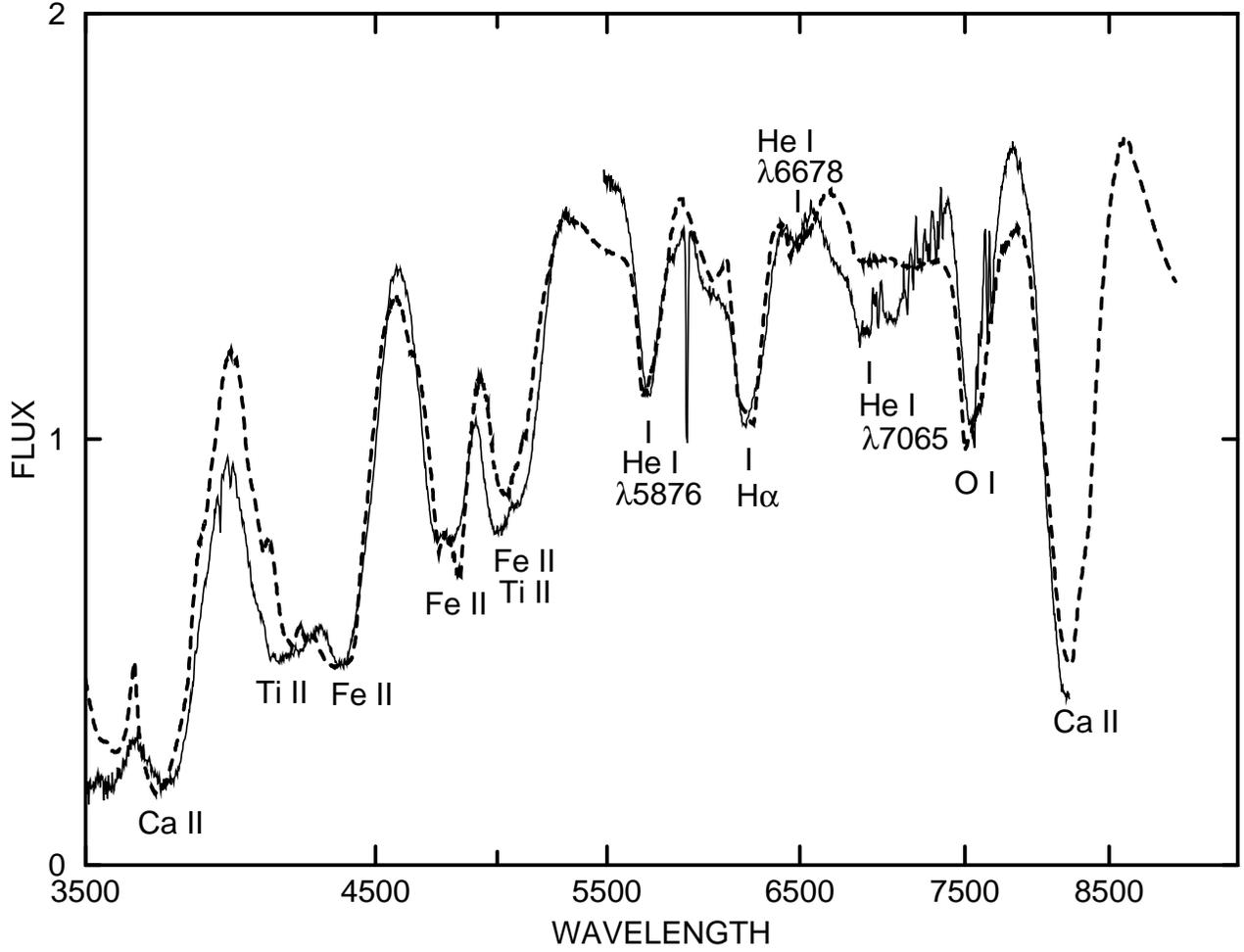}
\caption{ The day~+4 spectrum of SN~1994I ({\sl solid line}) is
compared to a {\bf Synow} synthetic spectrum ({\sl dashed line}) that
has \vphot = 11,000 \kms\ and includes lines of hydrogen, He~I, O~I,
Mg~II, Ca~II, Fe~II, and Ti~II.  }
\end{figure}

\begin{figure}
\includegraphics[width=.8\textwidth,angle=270]{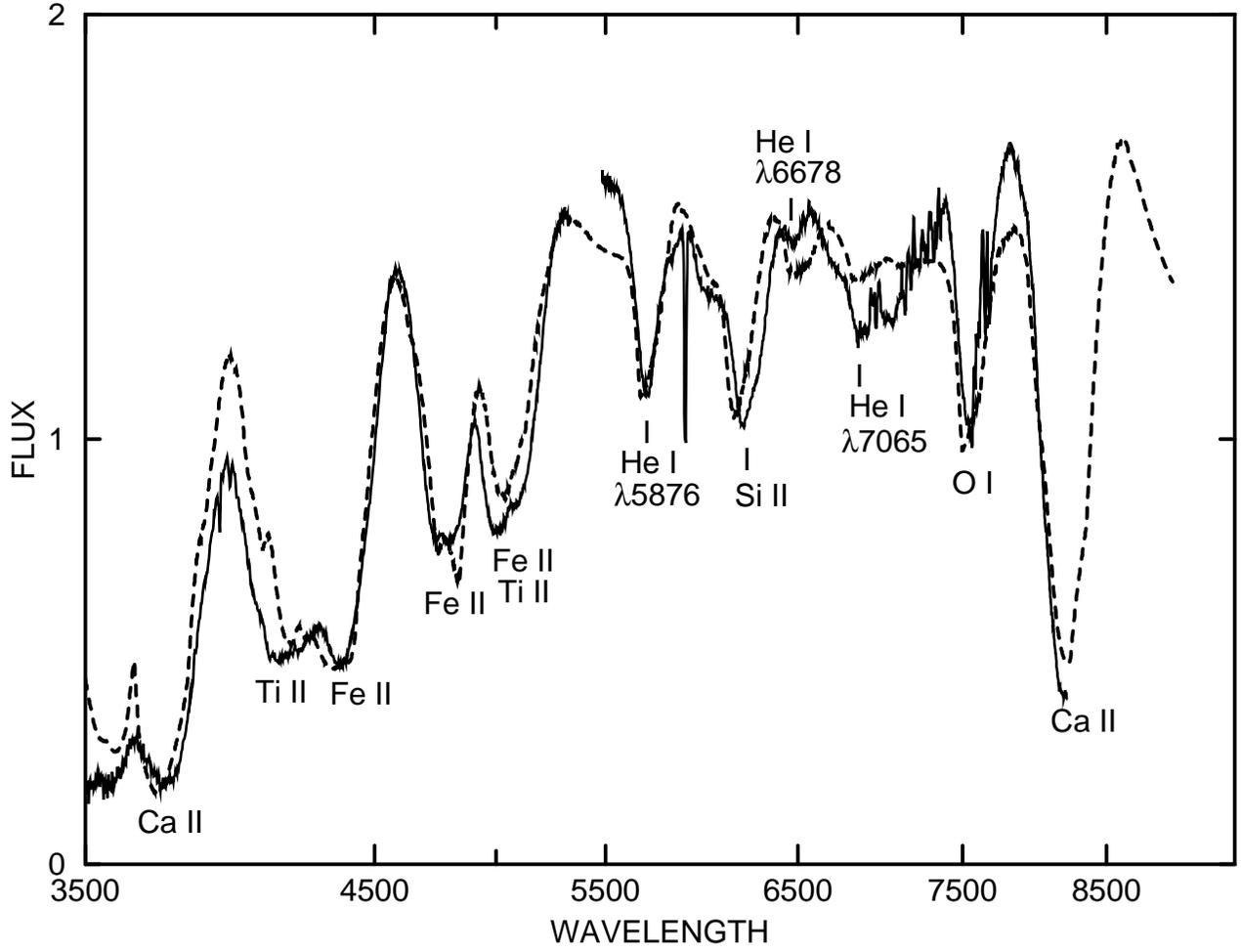}
\caption{Like Figure~5 except that the synthetic spectrum includes
  lines of Si~II instead of hydrogen. }
\end{figure}

\begin{figure}
\includegraphics[width=.8\textwidth,angle=0]{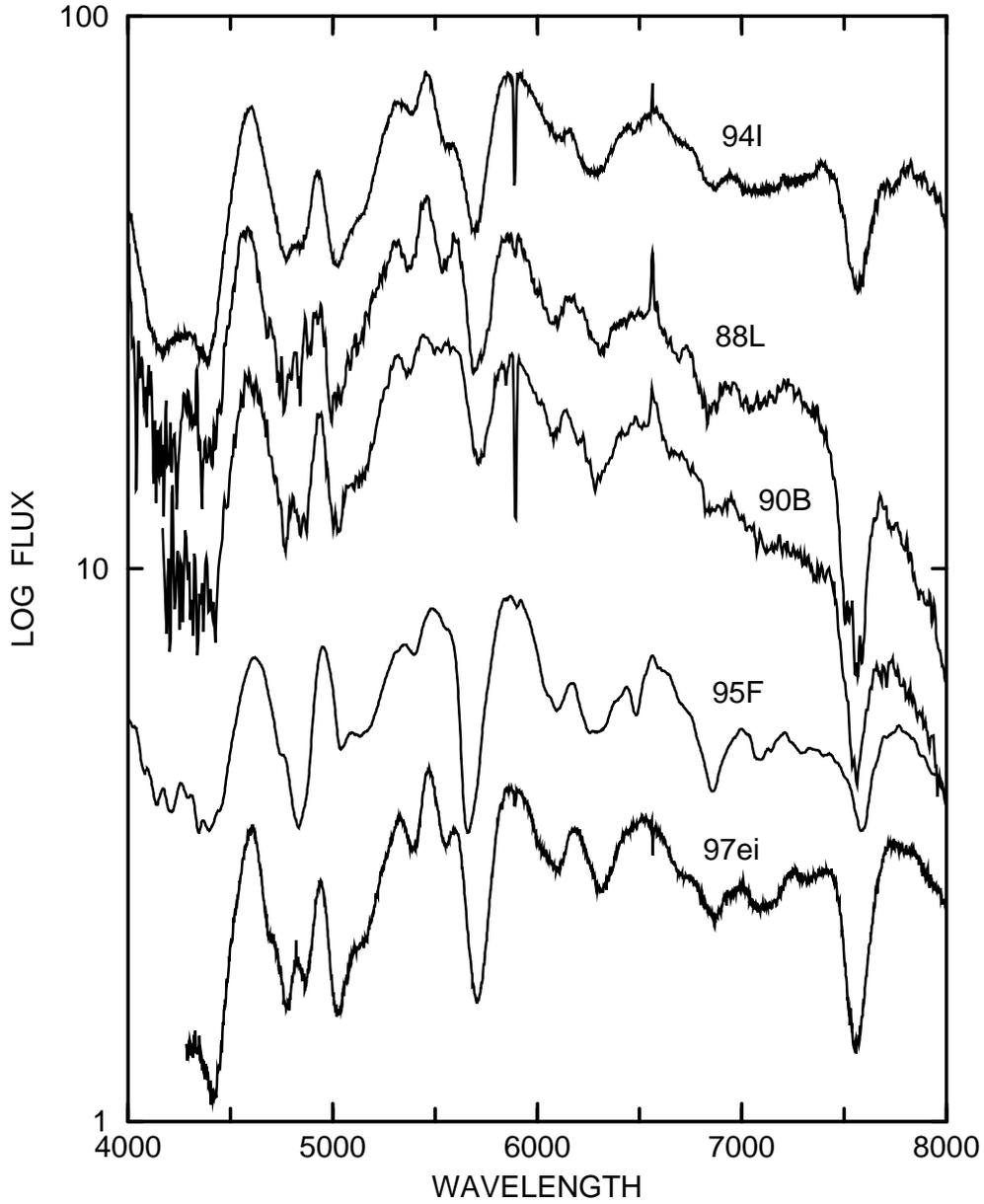}
\caption{A day~+10 spectrum of SN~1994I is compared with spectra of
 four other SNe~Ic (from Matheson et~al. 2001).  The flux scale is
 arbitrary and the spectra have been tilted to facilitate the
 comparison of spectral features.  The spectrum of SN~1990B was
 obtained on day~+5.  The spectra of SN~1988L, SN~1995F, and SN~1997ei
 were obtained 11, 10, and 6 days after discovery, respectively; the
 dates of $B$ maximum are unknown.}
\end{figure}

\clearpage

\begin{deluxetable}{lccccc}
\tablenum{1}
\setlength{\tabcolsep}{4pt}

\tablecaption{{\bf Synow} Fitting Parameters for Figures 4 and 5}

\tablehead{ \colhead{ } & \colhead{\lam$_{\rm ref}$} & \colhead{$\tau$} &
\colhead{$v_e$} & \colhead{$\tau$} & \colhead{$v_e$}\\

\colhead{ } & \colhead{ } & \colhead{Fig.~4} & \colhead{Fig.~4} &
\colhead{Fig.~5} & \colhead{Fig.~5} }

\startdata

\halpha & \lam6563 & 0.5 & 20 & 0.4 & 20\\

He~I & \lam5876 & 2.5 & 1 & 1.2 & 1\\ 

O I & \lam7773 & 1 & 1 & 2 & 1\\

Mg~II & \lam4481 & 2 & 1 & 2 & 1\\

Ca~II & \lam3934 & 80 & 4 & 80 & 4\\

Ti~II & \lam4550 & 0 & & 5 & 1\\

Fe~II & \lam5018 & 15 & 2 & 4 & 2\\

\enddata
\end{deluxetable}


\clearpage

\begin{references}

\reference{} Anupama, G. C., Sahu, D. K., Deng, J., Nomoto, K.,
Tominaga,~N., Tanaka,~M., Mazzali,~P.~A., \& Prabhu,~T.P. 2005, ApJ,
631, L125

\reference{} Baron, E., Branch, D., Hauschildt, P. H,
Filippenko,~A.~V., \& Kirshner,~R.~P. 1999, ApJ, 527, 739

\reference{} Benetti, S., et~al. 2002, MNRAS, 336, 91

\reference{} Branch, D. 2001, in Supernovae and Gamma--Ray Bursts: The
Largest Explosions in the Universe, ed. M.~Livio (Cambridge: Cambridge
Univ. Press), 96

\reference{} Branch, D. 2002, in A Massive Star Odyssey: from Main
Sequence to Supernova, Proc. IAU Symposium No. 212, ed. K.~A.~van der
Hucht, A.~Herraro, \& C.~Esteban (San Francisco: ASP), 346

\reference{} Branch, D., et~al. 2002, ApJ, 566, 1005

\reference{} Deng, J. S., Qiu, Y. L., Hu, J. Y., Hatano, K., \&
Branch, D. 2000, ApJ, 540, 452

\reference{} Elmhamdi, A., Danziger, I. J., Branch, D., Leibundgut,
B., Baron,~E., \& Kirshner,~R.~P. 2006, A\&A, in press

\reference{} Filippenko, A. V. 1988, AJ, 96, 1941

\reference{} Filippenko, A. V. 1992, ApJ, 384, L37

\reference{} Filippenko, A. V. 1997, ARAA, 35, 309

\reference{} Filippenko, A. V., Matheson, T., \& Ho, L. C. 1993, ApJ,
415, L103

\reference{} Filippenko, A. V., Porter, A., \& Sargent, W. L. W. 1990,
AJ, 100, 1575

\reference{} Filippenko, A. V., et~al. 1995, ApJ, 450, L11

\reference{} Folatelli, et~al. 2006, ApJ, in press

\reference{} Foley, R. J., et~al. 2003, PASP, 115, 1220

\reference{} Fryer, C. L., \& Heger, A. 2005, ApJ, 623, 302

\reference{} Hamuy, M., et al. 2002, AJ, 124, 417

\reference{} Hamuy, M., \& Phillips, M. M. 1999, IAU Circ. 7310

\reference{} Hatano, K., Branch, D., Fisher, A., Millard,~J., \&
Baron,~E. 1999, ApJS, 121, 233

\reference{} Iwamoto, K., Nomoto, K., H\"oflich, P., Yamaoka,~H.,
Kumagai,~S., \& Shigeyama,~T. 1994, ApJ, 437, L115

\reference{} Iwamoto, K., et~al. 1998, Nature, 395, 672

\reference{} Jeffery, D. J., Branch, D., Filippenko, A. V., \&
Nomoto,~K. 1991, ApJ, 317, 717

\reference{} Lucy, L. B. 1991, ApJ, 383, 308

\reference{} Matheson, T., Filippenko, A. V., Li, W., Leonard,~W.~C.,
\& Shields,~J.~C. 2001, AJ, 121, 1648

\reference{} McFadyen, A. I., \& Woosley, S. E. 1999, ApJ, 524, 262

\reference{} Mazzali, P. A., Iwamoto, K., \& Nomoto, K. 2000, ApJ,
545, 407

\reference{} Mazzali, P. A., et~al. 2002, ApJ, 572, 61

\reference{} Mazzali, P. A., Deng, J., Maeda, K., Nomoto, K.,
Filippenko,~A.~V., \& Matheson, T. 2004, ApJ, 614, 858

\reference{} Millard, J., et~al. 1999, ApJ, 527, 746

\reference{} Nomoto, K., Yamaoka, H., Pols, O. R., van den
Heuvul,~E.~P.~J., Iwamoto,~K., Kumagai,~S., \& Shigeyama,~T. 1994,
Nature, 371, 227

\reference{} Petrovic, J., Langer, L., Yoon, S.-C., \& Heger,~A. 2005,
A\&A, 435, 247

\reference{} Podsiadlowski, P. 1996, in Hydrogen Deficient Stars,
ed. C.~S.~Jeffery \& U.~Heber (San Francisco: ASP), 419

\reference{} Tominaga, N., et al. 2005, ApJ, 633, 97

\reference{} Wheeler, J. C., \& Harkness, R. P. 1986, in Galaxy
Distances and Daviations from Universal Expansion, ed. B.~F.~Madore,
R.~B.~Tully (Dordrecht: Reidel), 45

\reference{} Wheeler, J. C., Harkness, R. P., Clocchiatti, A.,
Benetti,~S., Brotherton,~M.~S., Depoy,~D.~L., \& Elias,~J. 1994, ApJ,
436, L135

\reference{} Woosley, S. E., \& Eastman, R. G. 1997, in Thermonuclear
Supernovae, ed. P~Ruiz--LaPuente, R.~Canal, \& J.~Isern (Dordrecht:
Kluwer), 821

\reference{} Woosley, S. E., \& Heger, A. 2006, ApJ, 637, 914

\reference{} Yoon, S.--C., \& Langer, N. 2005, A\&A, 443, 643

\end{references}
\end{document}